\documentstyle[preprint,aps]{revtex}
\newcommand{\nr}{I\!\!R}
\newcommand{\nn}{I\!\!N}
\newcommand{\nl}{I\!\!L}
\newcommand{\nm}{I\!\!M}

\begin{document}

\title{Higher-Dimensional Integrable Systems\\ from \\Multilinear
Evolution Equations}
\author{Jens Hoppe \footnote{Heisenberg Fellow. On leave of absence
from Karlsruhe University}\\
Institut f\"ur Theoretische Physik\\ETH H\"onggerberg\\CH 8093 Z\"urich}

\maketitle

\begin{abstract}
A multilinear $M$-dimensional generalization of Lax pairs is introduced
and its explicit form is given for the recently discovered class of
time-harmonic, integrable,
hypersurface motions in $\nr^{M+1}$.
\end{abstract}

\vspace{2cm}

\centerline{\bf ETH-TH/96-04}
\newpage

In \cite{1} the explicit form of a triple $(L,M_1,M_2)$, depending on
2 spectral parameters and 4 time-dependent functions
$x_i(t,\varphi^1,\varphi^2,\varphi^3)$ from a 3-dimensional Riemannian
manifold $\sum$ to $\nr$ was given such that (with $\rho$ a non-dynamical
density on $\Sigma$)
\begin{equation}
\label{eq:1}
 \dot{L}=\frac{1}{\rho}\in_{rsu}\frac{\partial L}{\partial\varphi^u}
   \frac{\partial M_1}{\partial\varphi^r}
    \frac{\partial M_2}{\partial\varphi^s}
\end{equation}
is equivalent to the equations
\begin{equation}
\label{eq:2}
 \dot{x}_i=\frac{1}{\rho}\in_{ii_1i_2i_3}\in_{r_1r_2r_3}
  \partial_{r_1} x_{i_1}\partial_{r_2}x_{i_2}\partial_{r_3}x_{i_3}\,,
\end{equation}
describing the integrable motion of a hypersurface $\hat{\Sigma}$ in
$\nr^4$ whose time-function (the time at which $\hat{\Sigma}$ reaches
a point $\bbox{x}\in\nr^4$) is harmonic \cite{2}. 

The purpose of this
note is to give the explicit generalization of this construction to
an arbitrary number of dimensions, $M(=\mbox{\rm dim}\, \Sigma$). Let
\begin{equation}
\label{eq:3}
 z_1=x_1+ix_2,\quad z_2=x_3+ix_4,\dots \quad.
\end{equation}
For {\bf even $M$} $(=2m)$ one may take
\begin{eqnarray}
\label{eq:4}
 L &=& \sum_{a=1}^m\left(\lambda_az_a-\frac{\bar{z}_a}{\lambda_a}\right)
       + 2\sqrt{m}\,x_N\nonumber\\
 M_a &=&\frac{i}{2}\left(\lambda_az_a+\frac{x_N}{\sqrt{m}}\right)
       \qquad a=1\dots m\nonumber\\
 M_{m+a'} &=&\left(\frac{4}{m}\right)^{\frac{1}{N-3}}
           \left(\frac{\bar{z}_{m+1-a'}}{\lambda_{m+1-a'}}-
            \frac{\bar{z}_{m-a'}}{\lambda_{m-a'}}\right)
       \qquad a'=1\dots m-1
\end{eqnarray}
depending on $m$ spectral parameters, $\lambda_a$, and $N=M+1$ functions
$x_i(t,\varphi^1,\dots,\varphi^M)$; letting
\begin{equation}
\label{eq:5}
 \{f_1,\dots,f_M\}:=\frac{1}{\rho(\varphi^1\dots\varphi^M)}
  \in_{r_1\dots r_M}\partial_{r_1}f_1\dots\partial_{r_M}f_M\,,
\end{equation}
\begin{equation}
\label{eq:6}
 \dot{L}=\{L,M_1,M_2,\dots,M_{2m-1}\}
\end{equation}
will then be equivalent to the equations of motion (as above, $\bar{z}_a$
denoting the complex conjugate of $z_a$)
\begin{eqnarray}
\label{eq:7}
 \dot{z}_a &=& -i\left(\frac{i}{2}\right)^{m-1}
         \{z_a,z_{a+1},\bar{z}_{a+1},\dots,z_{a-1},\bar{z}_{a-1}\}
\nonumber\\
\dot{x}_N &=& \left(\frac{i}{2}\right)^m
   \{z_1,\bar{z}_1,\dots,z_m,\bar{z}_m\}\,.
\end{eqnarray}
For {\bf odd $M$} $(=2m+1)$, rather than given a particular form of
$L,M_1,\dots,M_{2m}$ that would make
\begin{equation}
\label{eq:8}
 \dot{L}=\{L,M_1,\dots,M_{2m}\}
\end{equation}
equivalent to the equations of motion
\begin{equation}
\label{eq:9}
\dot{z}_\alpha=-i\left(\frac{i}{2}\right)^m
  \{z_\alpha,z_{\alpha+1},\bar{z}_{\alpha+1},
   \dots,z_{\alpha-1},\bar{z}_{\alpha-1}\}\,,\quad \alpha=1,\dots,n=m+1
 \quad,
\end{equation}
let me in this case stress the simple general nature of
the construction: Think of
\begin{equation}
\label{eq:10}
 L=L_1\lambda_1z_1+L_2\frac{\bar{z}_1}{\lambda_1}+\dots+L_{N-1-}\lambda_m
z_m+L_N\frac{\bar{z}_m}{\lambda_m}\quad,
\end{equation}
and likewise $M_1,\dots,M_{2m}$, as $N=2n$ dimensional vectors
$\bbox{L},\bbox{M}_1,\dots,\bbox{M}_{2m}$ in a vectorspace $V$ with
basis $\lambda_1z_1,\dots,\frac{\bar{z}_m}{\lambda_m}$. The wanted
equivalence of (\ref{eq:8}) with (\ref{eq:9}) may then be stated
as the requirement that
\begin{equation}
\label{eq:11}
 \det\left(\bbox{L},\bbox{M}_1\bbox{M }_2\dots\bbox{M}_{2m}
\bbox{e}_j\right)=i\left(\frac{2}{i}\right)^m\hat{\bbox{L}}\cdot\bbox{e}_j
\quad,
\end{equation}
where $\bbox{e}_j=(0\dots010\dots 0)^{tr}$ and
\begin{equation}
\label{eq:12}
\hat{\bbox{L}}=\left(L_2,L_1,L_4,L_3,\dots,L_N,L_{N-1}\right)\,.
\end{equation}
Multiplying (\ref{eq:11}) with the $j$-th component of $\bbox{L}$
(or any of the $\bbox{M}$'s), and summing over $j$, one finds that all
$2m+1$ vectors $\bbox{L},\bbox{M}_1,\dots,\bbox{M}_{2m}$ have to be
perpendicular to $\hat{\bbox{L}}$; in particular,
\begin{equation}
\label{eq:13}
\hat{\bbox{L}}\cdot\bbox{L}=2\left(L_1L_2+\dots+L_{N-1}L_N\right)=0\,.
\end{equation}
Choosing $\bbox{M}_1,\dots,\bbox{M}_{2m}$ to be also perpendicular
to $\bbox{L}$, the only remaining condition,
obtained by multiplying (\ref{eq:11}) by $\hat{L}_j$ 
(and summing), becomes
($\sim$ denoting the projection onto the $2n-2=2m$-dimensional
orthogonal complement of the $\bbox{L},\hat{\bbox{L}}$-plane)
\begin{equation}
\label{eq:14}
 \det \left(\tilde{\bbox{M}}_1,\dots\tilde{\bbox{M}}_{2m}\right)=i
  \left(\frac{2}{i}\right)^m\,,
\end{equation}
\nopagebreak{which exhibits the large freedom in choosing the $\bbox{M}$'s
(for fixed $\bbox{L}$).} A similar reasoning applies directly to the real
equations (cp. \cite{2}),
\begin{equation}
\label{eq:15}
 \dot{x}_i=\frac{1}{M!}\in_{ii_1\dots i_M}\{x_{i_1},\dots,x_{i_M}\}\,;
\end{equation}
the Ansatz $L=\sum_{i=1}^N\nl_ix_i,M_1=\sum \nm_{1i}x_i,\dots$ immediately
implies
\begin{equation}
\label{eq:16}
\sum^N_{i=1}\nl_i^2=0\,,
\end{equation}
making $L^l,l\in\nn$, a harmonic polynomial of $x_1,\dots,x_N$ (while its
integral is time-independent, due to (\ref{eq:6}) resp. (\ref{eq:8})) no
matter whether $m$ is odd, or even.

\end{document}